\documentclass[11pt]{article}
\renewcommand{\thefootnote}{\fnsymbol{footnote}}
\def\etal{\hbox{\it et al.}{}}
\def\ibid#1#2#3{{\hbox{\it ibid.}{}}{\bf #1} (#2) #3}
\def\eg{\hbox{\it e.g.}{}}
\def\etc{\hbox{\it etc.}{}}
\def\ie{\hbox{\it i.e.}{}}
\def\nn{\hspace{2mm}}
\def\sss{\scriptscriptstyle}

\newcommand{\GeV}{\mbox{\rm GeV}}

\def\sVEV#1{\left\langle #1\right\rangle}

\def\abs#1{\left| #1\right|} 

\begin{document}
\begin{titlepage}
\hfill
\vbox{
    \halign{#\hfil        \cr
           GUTPA/01/04/01 \cr
           NBI-HE-01-06    \cr
           hep-ph/0104161 \cr
           } 
      }  
\vspace*{20mm}
\begin{center}
{\Large {\bf  Standard Model Higgs Boson Mass from Borderline  %
Metastability of the Vacuum}\\}

\vspace*{15mm}
{\ C. D. Froggatt}$^{\it a,}$\footnote[3]{E-mail: c.froggatt@physics.gla.ac.uk},
{\ H. B. Nielsen}$^{\it b,}$\footnote[4]{E-mail: hbech@alf.nbi.dk}
and {\ Y. Takanishi}$^{\it b,}$\footnote[5]{E-mail: yasutaka@nbi.dk}

\vspace*{1cm} 
{\it $^a$ Department of Physics and Astronomy,\\
Glasgow University,\\
Glasgow G12 8QQ, Scotland}\\
\vskip .5cm
{\it $^b$The Niels Bohr Institute,\\
Blegdamsvej 17, DK-2100 Copenhagen {\O}, Denmark}\\

\vspace*{.5cm}
\end{center}

\begin{abstract}
We have studied imposing the condition that the Standard Model 
effective Higgs potential should have two approximately degenerate 
vacua, such that the vacuum we live in is just barely metastable:  
the one in which we live has a
vacuum expectation value of $246~\GeV$ and the other one 
should have a vacuum expectation value of order the Planck scale. Alone 
borderline metastability gives, using the experimental top 
quark mass $173.1\pm4.6~\GeV$, the Higgs mass prediction 
$121.8\pm11~\GeV$. The requirement that the second minimum be at the 
Planck scale already gave the prediction $173\pm4~\GeV$ 
for the top quark mass according to our 1995 paper.


\vskip 3mm 
\noindent\ 
\begin{tabular}{ll}
$\!\!\!$Keywords:& $\!\!\!$Standard Model Higgs mass, Top quark mass, 
Metastable vacuum,\\ 
& $\!\!\!$Multiple Point Principle
\end{tabular}
\end{abstract}
\vskip 12mm

April 2001

\end{titlepage}
\renewcommand{\thefootnote}{\arabic{footnote}}
\setcounter{footnote}{0}
\setcounter{page}{2}
\newpage
\section{Introduction}

The four LEP collaborations, ALEPH~\cite{ALEPH}, 
DELPHI~\cite{DELPHI}, L3~\cite{L3} and OPAL~\cite{OPAL}, have
recently reported on the search for the Standard Model (SM) 
Higgs boson using data collected during 1999-2000. The 
combined results are consistent
with the hypothesis of the production of the SM Higgs boson
with a mass around $115~\GeV$, and an observed excess in 
the combined data set  
of $2.9~\sigma$~\cite{Igo}. Of course this experimental signal 
does not give statistically safe evidence for the existence of 
a Higgs boson with this mass. However if it should turn out 
that our calculation together with the eventual improved 
accuracy in the top quark mass measurements finally gives a 
Higgs mass of $115~\GeV$ rather 
precisely: then that would not only support our model but also the 
existence of a real Higgs boson causing the LEP events.   

Two of us~\cite{FN} have already 
predicted the SM Higgs boson mass using the philosophy of the 
Multiple Point Principle~\cite{MPP1,MPP2} (MPP) and also predicted 
the top quark mass using one more assumption. 
It has been argued that a mild form of locality breaking 
in quantum gravity due to baby universes~\cite{baby}, say, which are expected 
to render coupling constants dynamical, leads to the 
realization of the MPP in Nature:
Nature should choose the coupling constants such that
several phases can coexist, $\ie$ the vacuum can exist 
in ``degenerate'' phases. The MPP implies that, with renormalisation
group corrections, the SM should have two minima in the Higgs field 
effective potential and the values at the two minima should be 
the same, $\ie$ 
$V_{\rm eff}(\phi_{\rm min,1})=V_{\rm eff}(\phi_{\rm min,2})$.
This condition tells us that the vacuum in which we live is stable: 
we are ``safe'' from the danger of vacuum decay, but lie on the 
borderline for stability.
The other assumption was that the second minimum, in 
which we do not live, has a Higgs field vacuum expectation 
value (VEV) of the order of the Planck scale, 
$\sVEV{|\phi_{\rm min,2}|} \approx M_{\rm Planck}$, $\ie$ we 
require a strong first-order phase transition between the two vacua
from the point of view of a fundamental scale identified 
with the  Planck scale. Using the usual renormalisation group methods,
these two assumptions lead to rather precise predictions for
the top (pole) quark mass, $M_t=173\pm4~\GeV$, and the Higgs 
boson mass, $M_H=135\pm9~\GeV$. Therefore one might 
say that the MPP prediction for the Higgs mass 
does ``agree'' with the combined LEP
experimental results within $2.3$ standard deviations.

It is the purpose of this paper to develop further the MPP ideas 
of~\cite{FN} and investigate whether they might be made to agree better 
with the LEP data. So let us start by arguing that, with a 
Higgs mass of $115~\GeV$ and our reluctance to postulate new 
physics, we are almost driven to MPP in some form:
The number of LEP candidate Higgs particle events of mass
$M_H\approx115~\GeV$ is in agreement with the SM cross section times 
branching ratio for a SM Higgs particle. It must be admitted
however that they are also consistent with SUSY models, but only 
because the lower bounds on the other ($\eg$ pseudo-scalar) 
Higgs masses imply that the SUSY Higgs production cross section 
is close to that of the SM. In this paper we shall assume that 
the LEP events correspond to the production of a SM Higgs 
particle. This means that, unless new physics 
appears below the Planck scale, the effective 
Higgs potential~\cite{Sher1} will have a second minimum 
falling below the value at its present minimum, in principle 
signalling the instability
of the vacuum in which we live! Then there are only the 
following three possibilities: %
\begin{itemize}
\item There is important new physics at the GUT scale 
($10^{16}~\GeV$) or below.
\item We have to stretch the errors in the computation that the
borderline for stability should be $135 \pm 9~\GeV$. Since 
the stretching should be minimal it would be suggested 
that the Higgs mass be after all very close to the 
borderline, $\ie$ it would support the MPP hypothesis of 
the previous work~\cite{FN} that 
the Higgs mass should just lie on the vacuum stability borderline.
\item We must accept that the vacuum is {\it only metastable}. 
If we could replace the absolute vacuum stability border by 
the metastability border, we could bring the allowed SM Higgs mass
down by about $10~\GeV$~\cite{EQ}. But also in this case 
the Higgs mass would have to be
close to the border, now the metastability border. 
\end{itemize}

In this work we take seriously the last possibility, and 
investigate the idea of introducing a modified MPP to predict the SM Higgs 
boson mass from the requirement that our vacuum has only barely 
survived early cosmology (call it metastability MPP). 
We shall ask if there should 
be reasons that MPP ideas might after all lead to metastability MPP. 
Further we shall ask, phenomenologically, whether such a metastability
MPP fits the data.

Really the main point of the present article is to argue for and 
present the Higgs and top quark mass relation for an assumption, which 
we could reasonably call meta-MPP: the various coupling 
constants and mass parameters in the Lagrangian density are adjusted 
in such a way that there exist several (in the SM just $2$) vacua --
having for instance different Higgs field VEVs --
so that one/some of them is/are just on the borderline of 
decay into and getting eaten up by the other vacuum/vacua 
in the early Big Bang -- or perhaps later.

Let us stress that in the present paper, for the purpose of 
the stability studies of the vacuum, we assume that the 
SM is in practice valid almost all the way up to the Planck 
scale\footnote{%
The new physics in MPP is about what the values of the SM 
parameters should be, so that it is strictly 
complementary to the SM, $\ie$ in no way violates its truth.}. 
In section $5$ we shall argue that, 
with either the MPP or the meta-MPP assumption, 
it is suggested that any new 
fields would to a large extent be decoupled. So the 
assumption that we can use the pure 
SM up to the Planck scale is likely to be effectively valid, 
even if it is not completely true in reality.

In the following section, we briefly review the main idea of 
degenerate vacua -- Multiple Point Principle -- and the results from 
our previous paper~\cite{FN}. In section 3 we discuss the reasons 
for believing that the Higgs mass should lie on the  
borderline of metastability for the vacuum. In section $4$ we 
extract the relation between the top quark and Higgs masses from 
an article by Espinosa and Quir{\'o}s~\cite{EQ} and present 
the results for these masses predicted assuming one of them is 
known from experiment.  
Some discussion goes into section $5$ and finally section $6$ 
contains our conclusions.


\section{``Old'' prediction of Higgs mass using Multiple Point Principle}
\indent

In our previous work~\cite{FN} we applied the MPP
assumption to the pure SM, by postulating that the Higgs 
effective potential $V_{\rm eff}$ has two minima in the radial 
direction of the Higgs field -- really it means two rings of 
minima in the Mexican hat -- and that these have the same energy 
density. 

The relative height in energy density of the two minima 
$V_{\rm eff}(\phi_{\rm min,1})$ and $V_{\rm eff}(\phi_{\rm min,2})$
depends on the SM parameters 
such as the Higgs mass. When the Higgs mass for say fixed 
$\phi_{\rm min,1}$ is lowered, the second minimum 
energy density $V_{\rm eff}(\phi_{\rm min,2})$ 
turns out to go down too. Thus, as soon as the Higgs mass 
is lowered infinitesimally from the value satisfying the MPP, 
our vacuum becomes formally unstable under decay into the 
minimum number $2$. So the prediction of MPP is that the parameters
of the SM must be such that they lie exactly on the 
vacuum stability curve.

In the previous paper we used the results on the vacuum stability 
curve from the articles~\cite{Sher2,AI,CEQ} to obtain the MPP 
prediction for the Higgs mass, taking the other SM parameters from 
experiment. Using present day data this prediction is 
$M_H = 135\pm9~\GeV$.
Adding one extra assumption -- namely that the second 
minimum should be at a value of the Higgs field VEV 
of the order of magnitude of 
the Planck energy -- led us to a surprisingly accurate 
prediction for the top quark (pole) mass 
of $M_t = 173\pm4~\GeV$.
This result was obtained~\cite{FN} by solving the renormalisation 
group equations numerically for the running of the Higgs self-coupling 
$\lambda$, the top quark Yukawa coupling 
constant and the three SM fine structure constants 
as functions of the energy scale. We then used the approximation 
of the renormalisation group improved potential, by inserting 
the running couplings into the classical Higgs potential with
the scale identified as the Higgs field strength $\phi$. 

{\it A priori} one would have expected that taking 
the VEV at the second minimum $\phi_{\rm min,2}$ to be 
only order of magnitude-wise
that of the Planck scale would leave a big uncertainty in the 
predicted top quark mass. However, by a remarkable coincidence, 
this uncertainty turns out to be  
smaller than the experimental uncertainty in the top 
quark mass. This remarkable 
accident is due to ${\it (1)}$ there being an approximate fixed point
behaviour~\cite{Ross}, and ${\it (2)}$ the top mass approaching 
this limit from below at 
an extremely slow rate. In fact a numerical fit to the results 
from our renormalisation group calculations shows that 
the deviation of the top quark mass from its approximate infra-red 
quasi-fixed point limit varies, approximately, as the 
inverse of the 42nd root of the VEV at the second minimum:
\begin{equation}
  \label{eq:42th}
M_t \approx M_{t, {\rm qfp}} %
- \frac{C}{\sqrt[{42}]{\sVEV{\abs{\phi_{\rm min, 2}}}}}\nn,
\end{equation}
where $ M_{t, {\rm qfp}}$ denotes the  
infra-red quasi-fixed point value and $C$ is a constant. 

\section{Why should the world be on the borderline  %
of vacuum meta\-stability?}
\indent

As the underlying reason for the MPP we could take it that for some 
mysterious reason: There has to be a physical realization of both 
minima over comparable amounts of space-time four volume.
This ``mysterious'' requirement is somewhat analogous to the 
requirement of a microcanonical ensemble, $\ie$~imposing fixed 
energy rather than a given temperature. Very often such a 
microcanonical ensemble is forced to contain more than one 
phase, for example both ice and water should very often be present 
if a fixed number of water molecules are given a specified 
energy; in a fixed volume even a vapour phase and thus a triple 
point can be provoked. 

Let us consider whether standard cosmology can lead to such a 
co-existence of two phases in space-time.
For the original version of MPP with exactly degenerate vacua, 
the early Universe comes out of the Big Bang in our low VEV 
vacuum, since it has more light particles and thus a lower 
\underline{free} energy density $F\!=\!U - ST$ than the 
second, $\ie$ high VEV, vacuum. 
If now the high VEV vacuum had a slightly smaller zero 
temperature energy density, it should in principle be 
possible for a bubble of this new vacuum -- a vacuum 
bomb one could say -- to be produced when the Universe 
has cooled, which would expand and produce a domain of  
high VEV vacuum. However this does not seem very likely, 
because the wall that must separate the two phases 
becomes so high in energy per unit area that it becomes 
practically impossible to make transitions when the 
temperature is no longer within a factor $10^2$ or $10^3$ 
from the Planck temperature. So we conclude that, 
if the Higgs mass is above the metastability border mass 
of $\sim122~\GeV$, the low VEV vacuum comes out of the Big 
Bang and never develops into the high VEV one.  

On the other hand, for Higgs masses below the metastability 
border, the high VEV vacuum is -- by definition -- produced in the 
early Universe. Since the high VEV vacuum  
has the lower energy density  in this case, it can of course 
never return to the low VEV one.
It seems that comparable amounts of four 
space-time volumes for the two vacua could only exist for a Higgs mass 
just very close to the metastability border. Assuming continuity
of the four space-time volume ratio as a function of 
the Higgs mass, comparable amounts of the two vacua do occur 
for a Higgs mass equal to the {\it metastability} bound.  
This scenario corresponds precisely to our new meta-MPP, 
as defined in the introduction. 

In summary our motivation for preferring the meta-MPP is the 
belief that both vacua should be realized somewhere or 
some time; because otherwise how could they both have any 
physical significance? But, as we have just argued, this
is essentially impossible to achieve except for a Higgs 
mass very close to the metastability border. It must though 
be admitted that it may be very difficult to estimate in 
advance how big a difference in the zero temperature 
vacuum density is needed, if humanity is ever to produce 
a vacuum bomb with future technology.


\section{Results}
\indent

Espinosa and Quir{\'o}s~\cite{EQ} have calculated the lower 
bound on the (pole) 
Higgs mass, requiring the vacuum we live in not to have already decayed 
in the Big Bang, and found the following numerical approximation 
 for the metastability bound:
\begin{eqnarray}
\label{eq:espqu}
M_H/\GeV  \geq \left[2.278-4.654\left(\alpha_s(M_Z) - 0.124\right)\right] %
(M_t/\GeV) - 277\nn,
\end{eqnarray}
which is valid for $60~\GeV<M_H<125~\GeV$. We take a theoretical 
uncertainty~\cite{newisidori} in this formula for $M_H$ of
$\pm4.6~\GeV$, corresponding to an uncertainty of $\pm2~\GeV$ 
in $M_t$. 

The present values reported by the Particle Data Group~\cite{PDG} 
$\alpha_s(M_Z) = 0.1185(20)$ and $M_t=174.3\pm 5.1~\GeV$, coming  
from the direct  
observation of top quark events, give rise via the meta-MPP 
to the Higgs mass prediction, 
$M_H = 124 \pm 13~\GeV$. If one instead used the indirect value of the 
top quark mass $168.2{+9.6 \atop -7.4}~\GeV$, coming from
the SM electroweak fit, our meta-MPP prediction of the 
Higgs mass becomes $M_H = 110.5{+23\atop -18}~\GeV$. However combining 
all the information to fit the top quark mass in the SM, 
the Particle Data Group gets $M_t = 172.9\pm4.6~\GeV$ assuming a 
fixed Higgs mass of $98{+57\atop -38}~\GeV$. Now every 
unit in the logarithm of the Higgs mass is correlated to 
increasing the indirectly measured top quark mass by $7.5~\GeV$. If we 
thus contemplate a $15~\GeV$ larger Higgs mass than the $100~\GeV$ 
mass used, the indirect top mass should be increased by $1.06~\GeV$. 

Since in the averaging of indirect and direct top masses, one uses 
the weight $5.1^2/(5.1^2 +9.6^2)=0.22$ for the indirect mass, 
such an increase by $1.06~\GeV$ shifts the average up by 
$1.06~\GeV \cdot 0.22 = 0.23~\GeV$. In other words  
the Particle Data Group fitted top quark mass value 
$M_t = 172.9\pm4.6~\GeV$ gets increased to 
$173.1\pm4.6~\GeV$,  which is thus the best value to
use.  Our corresponding best 
present prediction for the Higgs mass is
\begin{equation}
\label{eq:higgsmass}
M_H= 121.8 \pm 11~\GeV\nn.
\end{equation}
If LEP has indeed determined the Higgs mass to be 
$M_H= 115{+1.3\atop -0.9}~\GeV$,
the uncertainty in its mass is
much less than the $\pm11~\GeV$ uncertainty
of our prediction. We should then rather use the LEP 
Higgs mass to predict the top quark mass:
\begin{equation}
   \label{eq:topmass}
   M_t = 170.2 \pm 2.0~\GeV\nn.
\end{equation}
\section{Discussion}
\indent

Since phenomenologically the meta-MPP seems to be in good agreement 
with data, we should like to present some further ideas 
supporting the two main ingredients in our calculation: 
the meta-MPP itself and essentially the pure SM being valid 
up to the Planck scale.

\subsection{Why MPP or meta-MPP?}

If one writes down just randomly a Lagrangian density, 
without caring for whether the Hamiltonian should be 
bounded from below, one would probably hit a Lagrangian 
density providing us with tachyons and negative 
$\lambda\,|\phi|^4$ terms so that there would be no bottom 
to the Hamiltonian at all!
So one might ask: How did it come about that there {\em is} a bottom -- 
seemingly at least -- in the energy density? An immediately 
suggestive answer would be that this has to happen otherwise the vacuum 
would decay forever. But if this kind of 
answer should be taken seriously, one should think that 
the cosmological development involved the decay of 
many candidate vacua 
until finally -- almost accidentally -- an 
approximate effective Hamiltonian with a bottom 
occurs. So this type of thinking would suggest 
that the Universe and the present vacuum may not 
at all have reached the true bottom of the energy density, but only 
an ``accidental'' metastable state on the way. 

An alternative answer to the question of the origin of 
the bottom in the energy density
could be that the form of the Lagrangian density, in 
terms of the parameters/couplings not already fixed 
by some symmetry, is such that there is a guaranteed 
bottom. The suggestive example of a symmetry set up 
along this line is SUSY. In SUSY there is automatically a 
guaranteed bottom in the energy density, 
because the 
Hamiltonian has the form $H=Q^\dagger\,Q$. 
The value of the energy density at the bottom is even zero, 
as is also the phenomenological cosmological 
constant to high accuracy. 

The cosmological constant being nearly zero suggests an idea 
about how to ``derive'' MPP:
Whatever is 
the reason for getting a bottom at zero energy density, as the 
cosmological constant problem suggests, it should have a 
high chance of giving several degenerate minima.
If it can give one zero, why not several of them? We are indebted to 
Susskind~\cite{susskind} for this argument supporting the MPP.
SUSY is actually an example of one of the models 
that can solve the 
cosmological constant problem, but only 
{\em if SUSY were an exact symmetry}. As 
we would generally expect, it actually usually predicts\footnote{However,
the MPP derived from SUSY of course does not give any interesting 
predictions for the values of the SUSY parameters.} the MPP. Namely, in 
SUSY, it is well-known that there are usually flat directions or 
several zero-energy density minima in the effective potential for the
scalars. 

\subsection{Why essentially the pure SM?}

The above calculations were made assuming that the SM 
really is valid all the way up to the Planck scale. 
This is perhaps not so realistic in the 
light of the evidence for finite neutrino masses, suggesting a 
``new'' scale at for instance the see-saw scale of around 
$10^{12}~\GeV$; or perhaps at some very low scale, but at least 
a new scale, neither Planck nor weak will do.
We can, however, give a very general -- but not perfectly 
functioning -- argument 
that the MPP, which is our main assumption, 
will tend to adjust the coupling constants so as to favour there being 
several separate ``sectors'' containing fields/particle types which 
{\em only interact weakly with the other sectors}. For this purpose we 
should make use of the following formulation of the MPP principle: %
The MPP (meta or full stability version) means that, by arranging 
the sizes of the various coupling constants and mass parameters in the 
Lagrangian density, there should, at least 
approximately, be made many degenerate minima (vacua). 
In the metastable case the minima are only approximately degenerate 
and transitions may take place between them but, by 
adjustment of the couplings \etc\, they are made to be just on the 
borderline for decay from one vacuum to another one.

The general argument from MPP in favour of there being approximately 
decoupled sectors goes like this: In order to get say $n$ 
degenerate vacua, there is a need for 
the fine-tuning of
$n-1$ coupling-parameters. We are not caring here for the cosmological 
constant problem that really all the vacua should have zero energy density,
but rather only ask for them all to have the same energy density.
Now imagine that there are some proposals for 
separable ``sectors'', in the sense that there 
are some groups of fields for which 
there are relatively few terms in the Lagrangian involving fields 
from both ``sectors''. Suppose, for instance, that there are two  
separable ``sectors'' and there are just $p$ interaction terms involving 
fields from both sectors. Suppose further that 
the MPP-machinery (whatever the physics behind
it may be) arranges the $p$ interaction terms between the sectors to be zero
and the now separate section $1$ to have $n_1$ (approximately) 
degenerate minima, while the
sector $2$ has $n_2$ degenerate minima. 
Then the whole theory will actually have gotten
$n_1\,n_2$ approximately degenerate vacua, because you have all the 
combinations of one sector $1$ vacuum with one sector $2$ 
vacuum. This arrangement would cost the fine-tuning of 
\begin{equation}
p + (n_1 - 1) + (n_2 -1)~{\rm parameters} \nn.
\end{equation}
If, however, the MPP-machinery does not choose the possibility of
decoupling by putting the $p$ interaction parameters to zero, then the
same number of approximately degenerate vacua, namely $n_1\,n_2$, would
cost the fine-tuning of 
\begin{equation}
n-1 = (n_1\, n_2 -1)~{\rm parameters} \nn.    
\end{equation}
If $p + (n_1 - 1) + (n_2 -1)$ is smaller
than $n_1\,n_2 -1$ it will pay better to 
decouple the two sectors. In that case one would therefore expect 
that the MPP will arrange the decoupling, in order to get the biggest number 
of degenerate vacua. 

Thus if new physics is sufficiently isolated, by there not being so
many interaction possibilities with say the SM particles,
then the MPP is expected to arrange the few left over interactions to get 
zero couplings. So we expect that the decoupling 
gets almost total, or rather as decoupled as can be made. 
This means that, even if there is some new physics, it is likely 
that the MPP will adjust the coupling constants so that the interaction
between the new physics particles and the SM ones is very often 
fine-tuned to zero. Therefore it is likely that the new 
physics can actually be safely ignored, in calculations involving the 
SM particles.
Especially the calculation of the Higgs mass $\etc$ constraints, 
from the requirement of degenerate minima in the SM, will
give very closely the same results as if one calculated it in the 
full model, provided the decoupling takes place as described.

An obvious example of such a separate sector, which may decouple 
approximately, would be one or more 
right-handed neutrinos getting their masses 
from a new Higgs field with its own characteristic mass scale, 
let us call it $\phi_{B-L}$ and assign it a gauged $B-L$ charge.
In this case, the
potentially separated sector consists of the Higgs field $\phi_{B-L}$,
one or more right handed neutrinos and a gauge field coupled to 
the $B-L$ charge. Since this gauge field couples to 
$B-L$, meaning the baryon number minus the lepton number, it will 
of course thereby be forced to interact also with those SM
particles which carry baryon number or lepton number.
But the SM (Weinberg-Salam) Higgs field,
$\phi = \phi_{\sss WS}$, 
does not carry any baryon or lepton number,
so a $B-L$ gauge field would only influence the SM 
effective Higgs potential {\em indirectly}.
However a coupling corresponding to the term  
$\lambda_{\rm int} |\phi_{\sss B-L}|^2 \,|\phi_{\sss WS}|^2$, 
causing a direct interaction between the two Higgs fields, is allowed
in the Lagrangian density. It will be the most important 
term for shifting the $n_1\,n_2$ minima away from their energy density 
values which would be obtained if the proposed ``see-saw'' sector and the 
SM sector were indeed decoupled. The above argument should therefore 
mean that the MPP will make the coupling constant 
$\lambda_{\rm int}$ for this interaction term very small or zero. 

So the see-saw scale will in first approximation not influence 
our calculation. However if there were a $B-L$ coupling gauge field, 
as alluded to, it would contribute to the running of the top quark 
Yukawa coupling constant $g_t$ making its beta function 
$\beta_{g_t}$ more negative in the range above the see-saw scale.
This would give a small correction to Eq.~(\ref{eq:espqu}), which 
deserves further investigation.


\section{Conclusion}
\indent

We have made what we can consider as a correction to the 
physical detail of the work by two of us in 1995 on the 
MPP (Multiple Point Principle) 
prediction of the SM Higgs mass.
The point is that we no longer consider it necessary that a 
principle roughly like the MPP  
should be valid in the 
strict sense of the exact degeneracy of the different minima.

Rather we now consider it more reasonable to assume a modified 
version, meta-MPP, with approximately the same conclusion as the exact 
degenerate vacuum energy density version of the MPP. The proposal 
for guiding us into this version of the MPP could be the requirement 
that, in the cosmological development of the 
Universe, four space-time regions having volumes 
of the same order of magnitude 
should be realized for the two different minima
in the SM effective Higgs potential; and with such a 
requirement there is only a good chance for realizing it if the coupling 
constants and the Higgs mass are adjusted to be on the 
\underline{meta}stability borderline. The old (exact degeneracy) MPP
would not realize the 
high Higgs field VEV vacuum at all, because the high 
temperature in the early Big Bang would ensure that the low VEV 
vacuum -- in which we live -- is  the
only one realized. Metastability MPP (meta-MPP) 
says that the Higgs mass is just on the borderline of 
Big Bang metastability for our vacuum (the $246~\GeV$ Higgs VEV 
minimum in the effective potential).
Our main point then is that, by replacing the 1995  
(exact degeneracy) version of the MPP by the metastability version,
we are led to a Higgs mass prediction of 
$M_H=121.8\pm11~\GeV$ in agreement (with $0.6~\sigma$ deviation) 
with LEP observations. 

We want to stress that one can consider our present 
prediction as a {\em correction} of the older one \cite{FN},  
$M_H\left|_{\rm old}=135\pm9~\GeV\right.$. The old MPP 
version suffers from the fact that the high Higgs VEV 
vacuum is very hard to realize. Speculative explanations for MPP 
have trouble in giving physical significance to a vacuum 
that is never realized. 

In the ``old'' work we also derived the top quark mass -- taking 
the other SM couplings from experiment -- by 
requiring the second minimum in the effective potential 
to occur for a Higgs field VEV of the Planck energy 
size -- order of magnitude-wise. In this way, we obtained 
the very successful top quark mass prediction of $173~\GeV$, with the 
surprisingly small uncertainty of the order of $\pm4~\GeV$. 
Preliminary estimates suggest that, when this prediction is 
corrected according to the modified (metastability) MPP, 
the value for the top quark mass is reduced by about 
$1~\GeV$. Such a reduction in the top quark mass would 
also cause about a $2~\GeV$ reduction in our $122~\GeV$ 
Higgs mass prediction (to $120~\GeV$). 

If the meta-MPP picture is correct, our predictions for the 
near future are:
\begin{itemize}
\item The LEP Higgs events will be confirmed.
\item The top quark mass will turn out to be on the low mass 
side but within one standard deviation of its present experimental 
value. Namely, with the present calculational accuracy and using the 
$115~\GeV$ LEP Higgs mass, we predict $M_t = 170.2 \pm 2~\GeV$.
\end{itemize}
  
In conclusion we would claim that our meta-MPP, which 
is after all a very simple principle, agrees rather well 
with the LEP candidate Higgs mass.

\smallskip

Note: In a recent preprint~\cite{newisidori} Isidori~$\etal$ give 
a discussion of the (meta) stability of the vacuum, in 
the light of the potentially found Higgs particle.
They neglect the effects of standard cosmology during 
the first second in the early Universe and  
concentrate on the possibility of vacuum  
decay via quantum tunnelling at a later epoch.

\section*{Acknowledgements}
We wish to thank M.~Quir{\'o}s, M.~Sher and L.~Susskind 
for useful discussions. 
One of us (Y.T.) wishes to thank A.~Blondel for a useful discussion 
of the LEP Higgs boson research. C.D.F. and H.B.N. thank the EU 
commission for 
grants SCI-0430-C (TSTS), CHRX-CT-94-0621, INTAS-RFBR-95-0567 
and INTAS 93-3316(ext). Y.T. thanks the Scandinavia-Japan 
Sasakawa foundation for grants No.00-22. 

\end{document}